\documentstyle[prl,aps,multicol,epsfig]{revtex}
\begin{document}
\draft
\title{SO(5) Symmetry in $t-$$J$ and Hubbard Models}
\author{W. Hanke, R. Eder, E. Arrigoni, A. Dorneich, S. Meixner, and M.~G. Zacher}
\address{$^1$Institut f\"ur Theoretische Physik, Universit\"at W\"urzburg,
Am Hubland,  97074 W\"urzburg, Germany}
\date{\today}
\maketitle

\begin{abstract}
Numerical and analytical results are reviewed, which support SO(5)
symmetry as a concept unifying superconductivity and antiferromagnetism in
the high-temperature superconductors. Exact cluster diagonalizations verify
that the low-energy states of the two-dimensional $t$$-$$J$ and Hubbard
mo\-dels, widely used microscopic models for the high-$T_{c}$ cuprates, form 
SO(5) symmetry multiplets. Apart from a small standard deviation ($\sim
J/10$), these multiplets become degenerate at a critical chemical potential 
$\mu _{c}$ (transition into doped system). As a consequence, the $d-$wave
superconducting states away from half-filling are obtained from the higher
spin states at half-filling through SO(5) rotations. Between one and two
dimensions, using weak-coupling renormalization, a rather general ladder
Hamiltonian inclu\-ding next-nearest-neighbor hopping 
can be shown to flow to an
SO(5) symmetric point. Experimental tests and consequences such as the
existence of a $\pi$-Goldstone mode both in the insulator and
superconductor and, in particular, the relationship between the
photoemission spectra of the insulator and superconductor, are 
emphasized.
\end{abstract}

\begin{multicols}{2}

\setlength{\textfloatsep}{2cm plus1cm minus0.5cm}
\setlength{\intextsep}{3cm plus1cm minus0.5cm}
\setlength{\floatsep}{10cm plus2cm minus.5cm}
\setlength{\dbltextfloatsep}{2cm plus1cm minus0.5cm}

\newcommand{\sof}{\hbox{$SO(5)$} }
\newcommand{\beq}{\begin{equation}}
\newcommand{\eeq}{\end{equation}}
\newcommand{\beqn}{\begin{eqnarray}}
\newcommand{\eeqn}{\end{eqnarray}}
\newcommand{\g}{\sigma}
\newcommand{\up}{\uparrow}
\newcommand{\down}{\downarrow}
\newcommand{\zz}[2]{\hbox{$Z_{#1#2}$} }
\newcommand{\dvfn}{\hbox{$\Delta v_F$} }
\newcommand{\sofm}{SO(5)}
\newcommand{\tp}{\hbox{$t_{\perp}$}}
\newcommand{\kx}{{k_{\parallel}}}
\newcommand{\ky}{{k_{\perp}}}
\newcommand{\kom}{\vv k}
\newcommand{\koms}{\vv k \g}
\newcommand{\vv}[1]{{\bf #1}}
\newcommand{\wtilde}{\widetilde}
\newcommand{\om}{\omega}
\newcommand{\lflow}{\omega}
\newcommand{\nonformale}[1]{#1}
\newcommand{\formale}[1]{}

\section{Introduction}
\label{intro}

As their most prominent universal feature, high-temperature superconductors
(HTSC) always display antiferromagnetism (AF) and $d-$wave superconductivity
in close proximity, in their phase diagram. Recently, a unifying theory has
been proposed, according to which these two at first sight radically
different phases are ``two faces of one and the same coin''. They are
unified by a common symmetry principle, the SO(5) symmetry \cite{Zhang}. In the
meantime, numerical simulations, i.e.~exact cluster diagonalizations, have
verified that this symmetry principle is obeyed for widely used microscopic
models for the high-$T_{c}$ cuprates, 
namely the (two-dimensional) $t$$-$$J$ and
Hubbard models, to within an accuracy which is better than the average
superconducting energy gap \cite{EdHaZh,MeHaDeZh,DeZhMeHa}.
These microscopic models are known to
reproduce salient experimental results of the normal state of the high-$T_c$
cuprates, including in particular the antiferromagnetic state. 
SO(5) symmetry then implies that these microscopic models also display $d-$%
wave superconductivity. This gives rise to a microscopic description of the
complex phase diagrams of the HTSC from the insulating antiferromagnetic
phase over to the metallic normal state and, finally, to the $d-$wave
superconducting phase. These results were further corroborated between one
and two dimensions by weak-coupling renormalization-group calculations,
which demonstrated that rather general ladder Hamiltonians, including
longer-ranged interactions \cite{ArHa,LinBalentsFisher} and hoppings
\cite{ArHa}, flow to an SO(5)-invariant fixed point \cite{ScZhHa}.
In this paper, the basic numerical and
analytical calculations are summarized, and recent insights regarding the
microscopic principle behind SO(5) symmetry \cite{EdDoZaHaZh} are reviewed.

The paper is organized as follows:

Section \ref{one_to_one} describes a new idea and 
dynamical principle behind SO(5)
symmetry. It is shown that a formulation in terms of triplet and hole
fluctuations around an ``RVB vacuum'' allows for a physically transparent
demonstration of the corner stone in SO(5) theory, i.e.~that AF and SC are
``two faces of one and the same coin''. By starting from this ``RVB
vacuum'' $|\Omega >$, which represents the spin liquid state 
at half-filling, we
demonstrate that an AF ordered state can be generated by forming a coherent
state 
\[
| \psi \rangle \sim e^{\lambda t_{z}^{\dag}\left( q=\pi \right) }
| \Omega \rangle , 
\]
which corresponds to $z$-like triplets condensed into the $q$$=$$\pi$, 
i.e.~AF wave-vector state.
However, in the SO(5) theory, the $z$-like triplet with momentum $\pi $
and the hole pair with momentum $0$ are components of one and the same
SO(5) vector. They are rotated into each other by the SO(5) generating
operator $\hat{\pi}$. This implies that the above coherent state with condensed
triplets can be SO(5)-rotated into a corresponding coherent state with 
$t_{z}^{\dag}(q=\pi )$ replaced by the (hole-) pair creation operator $\Delta
^{\dag}(q=0).$ This state corresponds to hole pairs condensed into the $q=0$
state, i.e.~a superconducting state. In other words: both the AF and the SC
state can be viewed as a kind of condensation out of the RVB state, or the
spin liquid. If the so-constructed AF state is the actual ground state at
half-filling, then this physically very appealing SO(5) construction
yields automatically the ground state in the doped situation, i.e.~the SC
state \cite{EdDoZaHaZh}.

In Section \ref{micro_models}, 
we summarize numerical evidence for the approximate SO(5)
symmetry of the two-dimensional (2D) Hubbard and $t$$-$$J$ model. The SO(5)
symmetry organizes the low-energy degrees of freedom and gives a new
microscopic picture of the transition from an 
AF ground state to the $d-$wave SC state
as the chemical potential is varied \cite{EdHaZh}: Our results show 
that the $d-$wave SC
ground states away from half-filling are obtained from the higher spin
states at half-filling through SO(5) rotations. We use a general and
direct recipe for checking microscopic Hamiltonians for SO(5) symmetry,
i.e.~the concept of ``superspin multiplets'' \cite{EdHaZh}. 
The basic idea is that the
low-energy excited eigenstates of a cluster display a definite structure
characteristic of a particular symmetry, a scheme which has already provided
convincing evidence for the long-range order of the Heisenberg AF on a
triangular lattice \cite{Bernu}. We have numerically diagonalized the low-lying
states of the $t$$-$$J$ model near half-filling and found that they fit into
irreducible representations (irreps) of the SO(5) symmetry group. At a
critical value $\mu _{c}$ of the chemical potential, the superspin
multiplets are nearly degenerate and, therefore, higher spin AF states at
half-filling can be freely rotated into dSC states away from half-filling.
Our overall exact-diagonalization results, when further combined with a
detailed spectroscopy of ``SO(5)-allowed'' and ``SO(5)-forbidden''
transitions between the superspin multiplets, suggest that the low-energy
dynamics of the $t$$-$$J$ model \cite{EdHaZh} can be described 
by a quantum nonlinear
$\sigma-$model with anisotropic couplings, and the transition is that of a
superspin flop transition \cite{Zhang}. It is truly remarkable that, while the
physical properties of AF and dSC states appear to be diagonal opposite, and
they are characterized by very different form of order, there exists,
nevertheless, a fundamental SO(5) symmetry that unifies them.

In these two-dimensional (2D) microscopic models, the undoped situation --
in agreement with the experimental situation in the cuprates -- corresponds
to that of a Mott insulator with broken SO(3) or spin rational symmetry:
long-range AF order is realized. The SO(5)-symmetry principle then tells
us how this long-range magnetic order and the accompanying low-energy spin
excitations are mapped onto the corresponding off-diagonal long-range SC
order and the low-energy ``Goldstone bosons'' (the $\pi -$mode) in the doped
situation \cite{Zhang,EdHaZh,MeHaDeZh}.

However, there exists also a second class of Mott-type insulators without
long-range AF order, i.e.~spin liquids, which have a gap to spin excitation.
There is growing experimental evidence that they are also intimately related
to the physics of high-$T_{c}$ compounds: Not only do these compounds show
above the N\'{e}el temperature and superconducting transition temperature at
small dopings signs of such a spin gap, but there exist also copper-oxides
with a CuO$_{2}$ plane containing line-defects, which result in ladder-like
arrangements of $Cu-$atoms (for a summary, see \cite{DagottoRice}). 
These systems can
be described in terms of coupled two-leg ladders \cite{DagottoRice}, 
which exhibit a spin
gap in the insulating compound and thus belong to the spin-liquid
Mott-insulator variety. Also the related ``stripe phases'' of the 2D
CuO$_{2}$ planes in the cuprate superconductors have recently received
considerable attention \cite{Tranquada,PoilblancRice}.
In these systems, the apparent connection
between the spin gap and superconductivity must be explained.

In order to illustrate how the SO(5) theory can, in principle, cope with
the challenge, an exactly SO(5) invariant ladder model has recently been
constructed \cite{ScZhHa}. In particular, it was shown that the spin-gap 
magnon mode
of the Mott insulator evolves continuously into the ``$\pi-$resonance'' mode
of the superconductor. This SO(5) symmetric model offers a reference point
around which departures from the SO(5) symmetry can be studied. Then,
two key questions arise, the first being the relationship of the
exact SO(5) ladder to the ``physical'' $t$$-$$J$ or Hubbard ladders and the
second regarding the connection to the other variety of Mott insulators,
i.e.~the ones with long-range AF order.

With regard to the first question, progress was recently made in the regime
of weak-coupling: Using the weak-coupling renormalization group method, two
independent works \cite{ArHa,LinBalentsFisher}
have demonstrated that rather generic ladder models at half-filling flow 
to an SO(5) symmetric fixed point. This work is reviewed in section 
\ref{ladders}.

In section \ref{ladders}, we also summarize work, which tries 
to numerically attack both
questions in the experimentally relevant intermediate to strong-coupling
regimes. As shown in this recent work \cite{EdDoZaHaZh}, 
SO(5) symmetry has profound
implications for the dynamical correlation functions, most notably the
single-particle spectrum. Specific predictions of SO(5), like a
``generalized rigid band behavior'' \cite{EdOhta,PrHaVdL}
 and the appearance of sidebands
in the inverse photoemission spectrum \cite{Dagotto} may indeed 
have been observed
long ago in the actual 2D $t$$-$$J$ model and also in recent angular-resolved
photoemission experiments \cite{Marshall}. Motivated by the present 
ladder theory, we
have carried out more detailed spectroscopies on the 2D model and obtained
results in strong support of SO(5).

Finally, we and other authors \cite{ArHa,LinBalentsFisher,EdDoZaHaZh,Duffy}
 have demonstrated that, despite the,
at first sight, rather unphysical parameter values of the SO(5) symmetric
ladder model, a ``Landau mapping'' to the more realistic $t$$-$$J$ model is
feasible. This may suggest that the SO(5) symmetric ladder is indeed the
generic effective Hamiltonian for 2-leg ladder systems and for the above
spin-liquid Mott-insulator variety, in general.

\section{One-to-one correspondence of antiferromagnetism and superconductivity}
\label{one_to_one}

In the HTSC, the dominant charge-carrier dynamics takes place in the
two-dimensional (2D) CuO$_{2}$-planes \cite{Marshall}. Each CuO$_{2}$ unit 
cell contains
an effective magnetic moment of spin $\frac{1}{2},$ essentially due to the
Cu ion. Neighboring Cu-spins form singlets -- the energy win due to the
singlet formation, the magnetic exchange $J$ is relatively large $\sim
120$meV $\sim 1400$K. On the other hand, the temperatures for the transition
into both low-temperature phases, the AF and the SC phases, $T_{N\acute{e}el}
$ and $T_{c}$, are both significantly lower and of similar magnitude $(\sim
250$K for $T_{N\acute{e}el}$ and $\sim 100$K for $T_{c})$. Already this
order of magnitude suggests that the mechanism of superconductivity does not
directly result from the singlet formation, but is instead related to the
mechanism, which results in AF in the undoped, insulating situation.
 
Indeed, both low temperature phases of the cuprates are ``ordered'': In the
undoped case, i.e.~in the insulator, we have AF order, in the doped case,
the phase coherence of the superconductor and both phases are in the HTSC in
immediate vicinity (the ``spin glass phase'', which
occurs in some cuprates, is likely due to disorder). Therefore, it seems
tempting to unify these low-temperature phases, despite the fact that on
first glance, they appear dramatically different: On the one hand, the
insulator and, on the other hand, the ideal conductor, i.e.~the
superconductor.

Let us consider, at first, the insulator: At high-temperatures $\sim 1000$ K,
the singlet pairs are completely disordered. This state is termed,
therefore, a spin liquid. How does one arrive from this disordered state at
high temperatures, at an ordered N\'{e}el state at low temperatures? To
solve this problem, we shall consider a dynamical
principle, which gives a particularly simple and transparent demonstration
of the key feature of SO(5) theory, namely the one-to-one correspondence
of antiferromagnetism and superconductivity: According to this dynamical
principle \cite{EdDoZaHaZh}, the ordered AF state can be considered
as a kind of Bose-Einstein
condensation of triplet excitations. The SC state, on the other hand,
corresponds to a Bose-Einstein type of condensation of Cooper pairs, namely
in the cuprate materials of hole pairs. Triplet excitation and hole-pairs
excitation are ``two phases of one and the same coin'' in similarity to other
unifying concepts such as the isospin theory of proton and neutron in nuclear 
physics \cite{Heisenberg}. The condensation energy yields then a new lower
temperature scale $T_{N\acute{e}el}$ and $T_{c}$.

Here, SO(5) symmetry enters in detail: Like a ``magnifying glass'', this
symmetry principle allows to differentiate the low-energy physics of the
order of $T_{c}$ and, in particular, to unify the, in principle, competing
AF and SC phases (specifically, this symmetry principle ``rotates'' triplet
excitations into hole-pair excitations and vice versa). Unifying principles
via symmetry are, of course, known from many fields of physics, such as
quarks in high-energy physics, the predictions of which were inspired by the
SU(3) classification of hadronic spectra. How does such a unification take
place in the HTSC, i.e.~in solid-state physics? 

The order parameter of the AF
is the sublattice magnetisation, a real 3-dimensional vector; if this vector
is different from zero, we have AF order. Consider now two fixed
neighboring sites in the 2-dimensional AF, let's say in the configuration ($%
\uparrow \downarrow $). This fixed spin configuration may be viewed as
being due to an superposition of the singlet ($\uparrow \downarrow
-\downarrow \uparrow $) with the $(S_{z}=0)-$triplet ($\uparrow \downarrow
+\downarrow \uparrow $). To create the macroscopic 2-dimensional AF in the
CuO$_{2}$ plane, therefore, we have to mix triplet excitations already at
high temperatures into all possible singlet configurations of the spin
liquid. The AF states then result as a kind of ``condensation'' of the
triplet excitations into the lowest possible energy state \cite{EdDoZaHaZh}.
The three
components of the triplet correspond to the three possible orientations of
the AF and the density of the ``condensed triplets'' corresponds to the
magnitude of the sublatice magnetisation. This dynamical principle
illustrates the relation between AF and SC states rather clearly: If, in the
AF state (we shall see below that it corresponds to a coherent state) the 
triplet excitation operator is replaced 
by a hole pair creation operator, we obtain
a coherent state which creates a macroscopic number of Cooper pairs, i.e.~a
SC state. The ``rotation'' AF$\rightarrow $SC, therefore, is described by an
operator, the $\hat{\pi}-$operator of SO(5) theory \cite{Zhang},
which replaces triplets by hole pairs.

In the following, we give a particularly simple illustration
for this key-feature of SO(5) theory, which can be worked out for a ladder
\cite{EdDoZaHaZh}, and discuss the equivalence of antiferromagnetism and
superconductivity for this example. 
The ground state of the ladder models is actually a
resonating valence bond (RVB) type of vacuum without
AF long-range-order \cite{Dagotto}. However, for
illustrative purposes, let us now construct an AF ordered
state (which is in general not an eigenstate of the
Hamiltonian) by condensing magnons into the RVB ground state. 
One can express the operator of staggered magnetization
in $z$-direction as 
\begin{equation}
M_s = \sum_n e^{i \pi n}
(\;P_n(\uparrow \downarrow) - P_n(\downarrow \uparrow)\;),
\end{equation}
where e.g.~$P_n(\uparrow \downarrow)$ projects onto states
where the $n^{th}$ rung has the configuration $\uparrow \downarrow$ 
(see Fig.~\ref{fig_ladder}).
It is easy to see that
\begin{eqnarray}
(\;P_n(\uparrow \downarrow) - P_n(\downarrow \uparrow)\;)
s_n^\dagger &=& t_{n,z}^\dagger, \nonumber \\
(\;P_n(\uparrow \downarrow) - P_n(\downarrow \uparrow)\;)
t_{n,a}^\dagger &=& \delta_{a,z}\; s_n^\dagger,
\end{eqnarray}
where $s_n^\dagger(t_{n,\alpha}^\dagger)$ creates a singlet($\alpha$-tiplet) 
on rung $n$. Therefore 
\begin{equation}
M_s = \sqrt{\frac{N}{2}}[\;
 t_z^\dagger(q=\pi) + t_z^{}(q=\pi)\;].
\end{equation}
If we now form the coherent state
\begin{equation}
|\Psi_\lambda \rangle =
\frac{1}{\sqrt{n}} e^{ \lambda \sqrt{N}
t_z^\dagger(q=\pi)} |\Omega\rangle,
\end{equation}
which corresponds to $z$-like triplets condensed into the
$k$$=$$\pi$ state, and
treat the ${\bf t}$ as ordinary Bosons, we obtain
\begin{equation}
\langle \Psi_\lambda | M_s | \Psi_\lambda \rangle =
\sqrt{2} \lambda N.
\end{equation}
This calculation shows that by starting from a spin liquid,
i.e.~an RVB vacuum,
an antiferromagnetically ordered state with $M_S$ in $z$-direction
can be generated by
condensing $z$-like triplet-excitations into the $k$$=$$\pi$ state.
At this point, we can invoke the SO(5) symmetry of the model
\cite{EdDoZaHaZh}, 
which tells us that since the  $z$-like triplet with momentum $\pi$ 
and the hole pair with momentum $0$
are two different components of a 5-vector, they are
dynamically indistinguishable. This means that the
AF state, with condensed triplets, can be SO(5)-rotated
into a state with condensed hole pairs.
It follows that, if the antiferromagnetic state were
the ground state at half-filling (which is the case for 2D materials
and physical ladder systems),
we can replace all $z$-like triplets by hole pairs
with momentum $0$ and by SO(5) symmetry automatically
obtain the ground state in the doped case. The latter then
consists of hole-pairs along the rungs condensed into the
$k$$=$$0$ states and thus is necessarily superconducting.
In other words: both the antiferromagnetic and the
superconducting state may be viewed as some kind of
condensate `on top of' the rung-RVB state.
SO(5) symmetry then simply implies that the
condensed objects are combined into a single
vector, whence the unification of antiferromagnetism and
superconductivity follows in a most natural way.

The above derivation makes sense only in a strong coupling limit,
where a ground-state description starting out from rung-singlets 
is appropriate.
One might expect, however, that similar considerations will apply
also for cases with a weak coupling within the 
rungs \cite{ArHa,LinBalentsFisher} (see also section \ref{ladders}).

In two dimensions (2D) the above interpretation of SO(5) symmetry
hinges crucially on two points: first, the excitation spectrum of a 
2D spin-liquid or RVB state must consist of Bosonic triplets
(rather than e.g.~Fermionic `spinons'), whose condensation 
into the minimum of their dispersion at $(\pi,\pi)$ leads to
antiferromagnetic ordering. This would imply that the antiferromagnetic
state in 2D also could be interpreted as a condensate of triplet Bosons.
Work along this line is in progress and promising \cite{Eder2}.

Second, these triplet-like Bosons with momentum $(\pi,\pi)$ must be dynamically
equivalent to a $d_{x^2-y^2}$ Cooper pair. Having established this equivalence 
one could immediately conclude that the antiferromagnetic phase 
(viz `condensate
of triplets') is identical to the $d$-wave superconducting phase 
(viz `condensate
of hole pairs') in the same sense as two nuclei belonging to the same 
Isomultiplet
are `identical'. Since the $\hat{\pi}$-operator in 2D \cite{Zhang} 
precisely converts a triplet 
with momentum $(\pi,\pi)$ into a $d$-wave hole pair 
(see section \ref{micro_models}),
the latter requirement 
is equivalent to $[H,\hat{\pi}]=\omega_0 \hat{\pi}$. 
This commutation relation, which will be discussed in detail in the 
next section,
tells us whether the dynamics of charge carriers respects SO(5) symmetry, 
i.e.~the Hamiltonian is in accord with the SO(5) rotation $\hat{\pi}$. 
Ideally, we should then
have zero for the commutator. This will happen for a critical chemical 
potential 
$\mu_c$, right where the transition into the doped system takes place. Further 
changing $\mu$ gives rise to a precession frequency, or energy to perform the 
AF $\rightarrow d-$SC rotation $\omega_0 \neq 0$.
The energy shift $\omega_0$ here would
correspond to the mass difference of proton and neutron in the Isospin algebra.
The validity of this eigenoperator relation has already been verified
numerically \cite{MeHaDeZh}.

\section{SO(5) Symmetry and Microscopic Models for HTSC}
\label{micro_models}

The dynamical principle of the previous section suggests to unify the
triplet excitations ${\bf t}^{\dagger }$ (corresponding to a
vector) and hole-pair excitations $\Delta $ (described by (Re$(\Delta )$), Im%
$(\Delta )$) into a $5-$dimensional vector, the so-called superspin vector.
More precisely, Zhang suggested to group the AF order parameter $%
{\bf S}({\bf Q})$ with ${\bf Q}=(\pi ,\pi )$
and $d$-wave SC order parameter $\Delta $ into a single five-component
vector, the superspin\cite{Zhang} 
\begin{equation}
n_{a} = \left( \Delta ^{\dag}+\Delta ,{\bf S}\left( {\bf Q%
}\right) ,-i\left( \Delta ^{\dag}-\Delta \right) \right) ,
\end{equation} 
where $\Delta=\left( i/2\right) \sum_{{\bf p}} 
(\cos p_x - \cos p_y) c_{{\bf p}}\sigma
_{y}c_{-{\bf p}}$ denotes the $d_{x^{2}-y^{2}}$ superconducting
order parameter. ${\bf S}\left( {\bf Q}\right) =\sum_{%
{\bf p}}c_{{\bf Q}+{\bf p}}^{\dag}%
{\bf \sigma }c_{{\bf p}}$ stands for the AF N\'{e}el
vector and $\sigma _{\alpha }$ are the Pauli spin matrices.
The transition
from AF to dSC is then viewed as a kind of ``superspin flop'' transition as
a function of the chemical potential or doping, where the direction of the
superspin changes abruptly. This transition or SO(5) rotation is formally
described by the so-called $\hat{\pi}-$operator.

The ``superspin flop'' transition is analogous to the problem of a spinning
top in a uniform gravitational field, or a magnetic moment in a uniform
magnetic field, as in nuclear magnetic resonance (NMR). As we shall see, in
the presence of the symmetry-breaking field (which, in our case is the
chemical potential or doping), the order parameter is forced to precess, and
this explicit breaking of the SO(5) invariance induces rotations between
AF and dSC states (with frequency $\omega_0$) and governs the competition 
between these two phenomena.

Mathematically, the $\hat{\pi}-$operator can straightforwardly be constructed
\cite{Zhang}: as a rotation from AF to dSC, it has to patch up the 
differences in the
quantum numbers of the corresponding order parameters. Since $\Delta $ has
spin $S$$=$$0$, whereas ${\bf S}\left( {\bf Q}\right) $
has $S$$=$$1$ (which is obvious from our triplet excitation picture in the last
section), $\hat{\pi}$ has to carry $S$$=$$1$, i.e.~it must be a triplet 
operator. $%
{\bf S}\left( {\bf Q}\right) $ has no charge, $\Delta $
has charge $\pm 2$, therefore, $\pi $ must create charge $\pm 2.$ Finally, $%
{\bf S}\left( {\bf Q}\right) $ has momentum $%
{\bf Q}=\left( \pi ,\pi \right) ,$ $\Delta $ instead has ${\bf Q}=0$,
thus, $\hat{\pi} $ must have momentum ${\bf Q}.$ Combination of these 
requirements fixes
the operator uniquely up to a form factor, which is given by $d$-wave
symmetry \cite{ra.ko.97}. This results in the $\hat{\pi} $ operators 
(we just give one of them) 
\begin{equation}
\label{def_pi+}
\hat{\pi} ^{\dagger}=\sum_{{\bf p}}(\cos p_{x}-\cos p_{y})c_{{\bf p%
}+{\bf Q}\uparrow }^{\dag}c_{-{\bf p}\uparrow }^{\dag}.
\end{equation}
Charge conjugation and spin lowering operation gives five other similar
operators. In real-space representation, the $\hat{\pi} \left( \hat{\pi} 
^{\dagger} \right) $
operator does precisely what we discussed in the previous section, namely it
replaces triplets oriented in $x$ and $y$ planar directions by $%
d_{x^{2}-y^{2}}$ hole (electron) pairs. These $\hat{\pi} $ operators were first
introduced by Demler and Zhang \cite{DemlerZhang} to explain the resonant 
neutron
scattering peaks in the $YBCO$ superconductors. Together with the total spin 
$S_{\alpha }$, which is the generator of the SO(3) spin rotation forming a
subgroup of SO(5), and the charge operator $Q$ generating the $U(1)$
charge symmetry (which is also a subgroup of SO(5)), the six $\hat{\pi}$
($\hat{\pi}^{\dag}$)
operators form the ten generators $L_{ab}$ of the SO(5) algebra (for
definition of $L_{ab}$, see \cite{Zhang} ). As constructed, the $\hat{\pi} $ 
operators
indeed rotate the AF order parameter into the dSC order parameter, i.e. 
\begin{equation}
\label{[pi+,S]}
\left[ \hat{\pi} _{\alpha }^{\dag},S\left( {\bf Q}\right) _{\beta }\right]
=i\delta _{\alpha \beta }\Delta ^{\dag},
\end{equation}
and vice versa.

We have, thus, apparently accomplished the task of unifying AF with SC: the
corresponding order parameters are grouped into a five dimensional object,
and SC is ``nothing'' but AF viewed in some rotated coordinates and vice
versa! This construction looks a bit similar to the unification of $%
{\bf E}$ and ${\bf B}$ by the Lorentz group. But, so
far, this is only a mathematical construction, we haven't asked if ``Mother
Nature'' approves the SO(5) construct or not.

In the high$-T_{c}$ problem, ``Mother Nature'' is very complicated, but we
can check the SO(5) symmetry within some microscopic models, which are
known to reproduce salient features of the phase diagram of HTSC. Such
microscopic Hamiltonians, i.e.~Hubbard and $t$$-$$J$ models, successfully model
the Mott-Hub\-bard insulator to metal transition
\cite{DemlerZhang,Anderson}, which is driven by
the Coulomb correlation $U\sim 10$eV, i.e.~by ``high-energy'' physics. They
also model prominent features of the magnetic interactions on an energy
scale of order $J\sim 0.1$eV 
\cite{EdOhta,PrHaVdL,Dagotto,DemlerZhang,Anderson,Scalapino,Schrieffer,Pines}. 
However, their low-energy
content of order of the average SC gap $\left( \sim J/5-J/10\right) $ has so
far eluded both analytical and numerical investigations \cite{Zhang}. We shall
demonstrate that SO(5) symmetry overcomes this major obstacle; it
clarifies the role of competing orders and gives a microscopic description
of the transition from AF to dSC ground states, as the chemical potential is
varied.

One can check the SO(5) symmetry by evaluating the commutator between the
Hamiltonian with the $\hat{\pi}-$operators. In particular, numerical
\cite{EdHaZh,MeHaDeZh} works,
summarized below, show that the $\hat{\pi} $ operators are approximate
eigenoperators of the Hubbard Hamiltonian, in the sense that 
\begin{equation}
\label{[H,pi+]}
\lbrack H,\hat{\pi} _{\alpha }^{\dagger}]=\omega _{0}\hat{\pi} _{\alpha }
^{\dagger},
\end{equation}
where the eigen-frequency $\omega _{0},$ which is the energy to perform the
AF$\rightarrow $dSC rotation, is of the order of $J$, and proportional to
the number of holes. This relation (\ref{[H,pi+]}) 
reminds us of the commutation
relation between the transverse spin components $S_{+}$ and $S_{-}$in a
magnetic field and the Zeeman Hamiltonian, with $\omega _{0}$ being
proportional to the ${\bf B}-$field. In our case, the SO(5)
symmetry is broken explicitly by the chemical potential, i.e.~$\omega _{0}$
is proportional to the hole count or doping. Thus, the pattern of explicit
symmetry breaking is simple and familiar, and, therefore, easy to
handle.

In the following, we employ a kind of ``computer spectroscopy'' to test
whether the dynamics of the charge carriers, i.e.~the Hamiltonian, respects
SO(5) symmetry and (\ref{[H,pi+]}) is fulfilled. Ever since the early days of
quantum dynamics, group theoretical interpretation of spectroscopy revealed
deep symmetry and profound unity of Nature. Atomic spectra can be fitted
into irreducible representations (irreps) of SO(3), and the regular
patterns which emerged from this classification offered fundamental
understanding of the periodic table. After the discovery of a large number
of hadrons, the ``embarrassment of riches'' was removed by the classification
of hadronic spectra into irreps of SU(3) and this hidden regularity
inspired the predictions of quarks, the fundamental building block of the
universe. In our quest for understanding the fundamental design of Nature,
the importance of symmetry can never be over-emphasized.

In our work, we used a different kind of spectroscopy and its classification
into a different kind of symmetry. The spectroscopy is performed on a
computer, which numerically diagonalizes microscopic Hamiltonians, 
i.e.~Hubbard and $t$$-$$J$ models widely believed \cite{DemlerZhang,Anderson}
to model high-$T_{c}$ superconductors.

As the simplest 2D lattice model for correlated electrons, the one-band
Hubbard model is defined as 
\begin{equation}
H=-t\sum_{\left\langle i,j\right\rangle ,\sigma }\left( c_{i\sigma
}^{\dag}c_{j\sigma }+\mbox{h.c.} \right)+U\sum n_{i\uparrow }n_{i\downarrow }, 
\end{equation}
with nearest-neighbor hopping $t$ and Coulomb correlation $U$. On the other
hand, the $t$$-$$J$ model has the Hamiltonian 
\begin{equation} 
H=P\left[ -t\sum_{\left\langle i,j\right\rangle, \sigma } \left( c_{i\sigma}
^{\dagger
}c_{j\sigma} + \mbox{h.c.} \right) +J\sum_{\left\langle i,j\right\rangle }
{\bf S}_i {\bf S}_j\right] P. 
\end{equation}
As in the Hubbard model, $\left\langle ij\right\rangle $ denotes a summation
over nearest-neighbors on the 2D square lattice and $P$ projects onto the
subspace with no doubly occupied sites. The latter constraint reflects the
strong correlations in the $U/t\rightarrow \infty $ limit of the Hubbard
model.

\begin{figure}
\vbox{%
\centerline{\epsfig{file=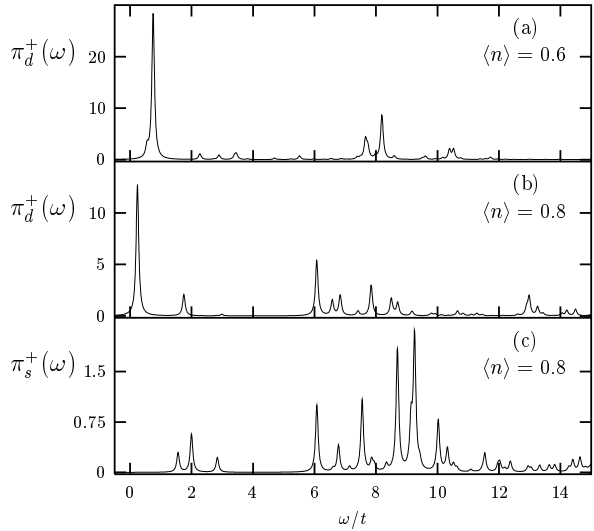,width=8cm}}
\narrowtext
\vspace{1mm}
\caption{\label{fig_10sites2D}Dynamic correlation functions of the $\sqrt{10} 
\times \sqrt{10}$ Hubbard model with $U$$=$$8t$: 
(a) $\pi_d^{\dag}(\omega)$-spectrum at $\langle n \rangle$$=$$0.6$, 
(b) $\pi_d^{\dag}(\omega)$-spectrum at $\langle n \rangle$$=$$0.8$, 
(c) $\pi_s^{\dag}(\omega)$-spectrum at $\langle n \rangle$$=$$0.8$}}
\end{figure}

First numerical evidence for the approximate SO(5) symmetry of the Hubbard
model came recently from exact diagonalizations of small-sized (10 sites)
clusters \cite{MeHaDeZh}, studying dynamic correlation functions involving 
the AF/dSC rotation $\hat{\pi} $ operator. 
We observe that (\ref{[H,pi+]}) is nothing but the ladder
operator relation familiar from standard Quantum Mechanics problems such as
the harmonic oscillator or the spin rising $(S_{+})$ and lowering $\left(
S_{-}\right) $ operators. If this equation is fulfilled, then the equal
level distance $\omega _{0}$ between the ``rungs'' of the ladder should
appear -- like in NMR or optical spectroscopy -- as a sharp peak in the 
$\hat{\pi}$$-$$\hat{\pi}$ correlation function, 
well separated from a higher-energy incoherent
background. This indeed is verified in Fig.~\ref{fig_10sites2D}(a),(b),
which displays a typical
result for the $\hat{\pi} _{d}$ correlation function in a ten-site 
Hubbard cluster
with $U/t=8$ for dopings of one hole-pair $\left( \left\langle
n\right\rangle =0.8\right) $ and two hole-pairs $\left( \left\langle
n\right\rangle =0.6\right)$ \cite{MeHaDeZh}. We note that $\omega _{0}$ 
is a small
energy, scaling with the hole count away from half-filling $\left(
\left\langle n\right\rangle =1\right) ,$ i.e.~$\omega _{0}\cong
J/2(1-\left\langle n\right\rangle )-2\mu .$

The approximate relation (\ref{[H,pi+]}) is highly nontrivial. One could ask 
if a similar relation would exist for a modified $\hat{\pi} $ operator which 
rotates AF into $s-$wave SC order parameters. The answer is negative 
\cite{EdHaZh,MeHaDeZh}, 
as seen in Fig.~\ref{fig_10sites2D}(c)
which demonstrates that a $\hat{\pi} -$rotation with $s-$wave symmetry
just generates an incoherent background and no sharp ``$\pi -$resonance''.
Therefore, there is only an approximate symmetry between AF and $d-$wave SC
(and not $s-$wave SC) near half filling.

In the next step, we use a most general and direct recipe for checking
microscopic Hamiltonians for SO(5) symmetry, i.e.~the concept of
``superspin multiplets'' \cite{EdHaZh}. We consider the $t$$-$$J$ model, 
which, because of
its more limited Hilbert space (no double occupancies), allows the exact
diagonalization of larger systems (18, 20 sites). Since the $t$$-$$J$ model
explicitly projects out the states in the upper Hubbard band, some of the
questions \cite{Greiter,Baskaran}
raised recently about the compatibility between the Mott
Hubbard gap and SO(5) symmetry can also be answered explicitly. In
particular, if there is an approximate SO(5) symmetry of the microscopic
model, the low-energy states of this model should fall into irreducible
representations (irreps) of SO(5). In a given quantum mechanical system,
the direction of the SO(5) superspin vector is quantized in a way similar
to an ordinary SO(3) spin, and the classically intuitive picture of the
precession of the SO(5) superspin vector under the influence of the
chemical potential \cite{Zhang} can be identified with the equal level-spacing
between the members of SO(5) multiplets carrying different charge.
Therefore, numerically identifying the low-lying states of the microscopic
model with the SO(5) irreps can lead to detailed understanding of the
one-to-one correspondence and the level crossing between the excited states
of the AF and the dSC states, and thereby lead us to the microscopic
mechanism by which the AF changes into the dSC state. While finite-size
calculations cannot generally be used to prove the existence of long-range
order in infinite systems, the spectroscopic information about the SO(5)
symmetry can be used as input for the effective field theory
\cite{Zhang,Burgess,Arovas} which
captures the low-energy and long-distance physics of the problem.

Exact diagonalizations (e.g.~\cite{Dagotto}) commonly study ground-state 
correlations,
but their spatial decay is often inconclusive as a test of order due to
small system size. Yet it is possible that the (excited) eigenstates show a
well-defined structure characteristic of a particular symmetry; this
provided the convincing evidence for long-range order in the spin-1/2
triangular lattice AF \cite{Bernu}. In our work, we have pursued exactly such a
program in exact diagonalizations of the $t$$-$$J$ model.

Consider as a simplest example the precession of a spin-$1/2$ system in a
homogeneous ${\bf B}-$field in $z-$direction. ${\bf B}$
breaks the SO(3) spin-rotation symmetry and the spin (expectation value)
precesses with Larmor frequency around the spatially fixed ${\bf B%
}-$direction. This can be read off in a spectroscopic experiment from the
multiplet structure of the possible spin states $\left( |\pm \frac{1}{2}%
>\right) ,$ the degeneracy of which is lifted by the ${\bf B}-$%
field (Zeeman effect). In this simplest case, the multiplet structure is
one-dimensional, extending in $S_{z}-$``direction''. 

\begin{figure*}
\vbox{%
\vspace{-4mm}
\centerline{\epsfig{file=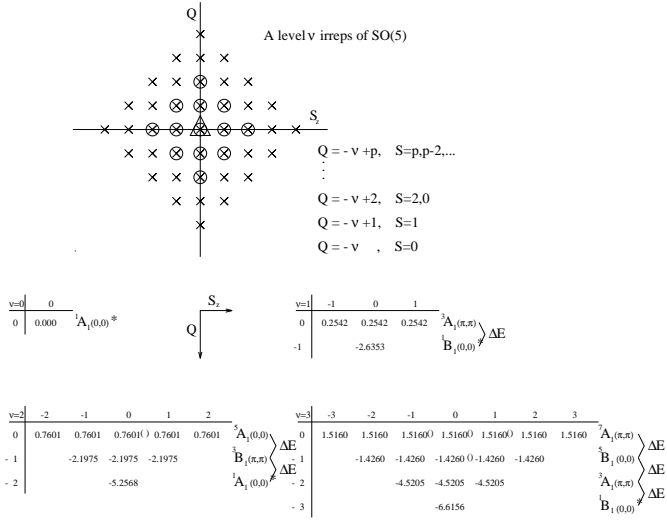,width=9cm}}
\narrowtext
\vspace{1mm}
\caption{\label{fig_irreps}The upper diagram illustrates general 
level $\nu$ irreps of
SO(5). Every state can be labeled by $Q$ and $S_z$. The maximal charge is
$Q=\pm \nu$. The states labeled by a $\times$ form the shape of a diamond,
while states inside the nested diamonds are labeled by $\circ$ and $\triangle$.
Overlapping states with same $Q$ and $S_z$ are distinguished by their 
$S$ quantum numbers. The lower diagrams are for $\nu=1,2,3$ irreps of SO(5).
The figure shows the energies of some low energy states for the $18$-site 
cluster with $J/t=0.5$. The states are grouped into different
multiplets and are labeled by the spin, point group symmetry,
and total momentum. $A_1$ denotes the totally symmetric,
$B_1$ the $d_{x^2-y^2}$-like representation of the
$C_{4v}$ symmetry group. The $( )$ symbol denotes
as yet unidentified members of the respective multiplet}}
\end{figure*}

In our ``computer spectroscopy'', we employ a formally analogous recipe for
checking the microscopic Hamiltonian for SO(5) symmetry:

In SO(5), the multiplet structure is two-dimensional, i.e.~spanning
both $S_{z}$-di\-rec\-tion and $Q$-di\-rec\-tion \cite{EdHaZh}. $Q=L_{15}$ is 
the total
charge, and, in our $t$$-$$J$ calculation, it stands for the number of doped
hole pairs, and, thus, for the transition from AF to dSC states. Formally,
the multiplets are constructed by observing that $\{S_{z},Q,C\}$ form a set
of commuting operators with their quantum numbers labeling states of an
SO(5) invariant Hamiltonian \cite{EdHaZh}. $S_{z}$ is the $z-$component of 
the total
spin $\left( S_{z}=-L_{23}\right) $ and $C$ the Casimir operator $%
\sum_{a<b}L_{ab}^{2}$, which is a natural generalization of the total spin
operator $\vec{S}^{2}$. Like ${\bf S}^{2}$ in the
familiar SO(3) spin-rotation symmetry, $C$ fixes the level $\nu $ of an
irreducible representation. Instead of $S(S+1)$ in SO(3), it takes the
value $\nu \left( \nu +3\right) $ for an SO(3) level $\nu $ irreps. Fig.
\ref{fig_irreps}
shows the first four $(\nu=0$ to $\nu=3)$ irreps of
SO(5) with the low-lying states of an 18-site $t$$-$$J$ model with $J/t=0.5$,
which is a typical $J$ value \cite{Dagotto}.

What is the physical meaning of these multiplets? In fact, there is a
definite physical meaning behind them, which is closely related to our
coherent-state description of the AF$\rightarrow $dSC transition in section
\ref{one_to_one}. 
The AF state is constructed from the linear superposition of the $Q=0,$ $%
S_{z}=0$ states in each of the level $\nu $ multiplets. At a given level $%
\nu $ this state contains $\nu $ magnons or triplet excitations, which
correspond to the $\nu -$th term in the power-series expansion of the
coherent-state operator, i.e.~$e^{\lambda t^{\dag}}=1+\lambda 
(t^{\dag})^{\nu=1}+\frac{\lambda ^{2}}{2}(t^{\dag})^{\nu=2}...$

Correspondingly, the dSC state is constructed from a linear superposition of
the lowest corner states of each level $\nu $ irreps, and may be viewed as
stemming from the power-series expansion, $e^{\lambda \Delta}=1+\lambda
\Delta ^{\nu =1}+\frac{\lambda ^{2}}{2}\Delta ^{\nu =2}...\smallskip $

From Fig.~\ref{fig_irreps} we note that the low-lying states indeed fit into 
the irreps 
of SO(5): all the different quantum numbers of the states are naturally
accounted for by the quantum numbers of the superspin. However, most
importantly, the levels with different charge $Q$ are nearly equally spaced!
This is explicitly indicated by the symbol $\Delta E$ in 
Fig.~\ref{fig_irreps}. More
precisely, the mean-level spacing within each multiplet (up to $Q=-2$) is $%
-2.9886$ with a standard deviation of $0.0769$. This standard
deviation is much smaller $\left(
\sim J/8\right) $ than the natural energy scale $J$ of the $t$$-$$J$ model and
comparable to or even smaller than the average SC gap. If one now adds the
chemical potential term, $H_{\mu} =-2\mu Q$, then at a chemical potential $\mu
_{c}$ comparable to the mean-level spacing, the superspin multiplets are
nearly degenerate. In other words, (each of the coherent-state contributions
to) the AF state can freely be rotated in (the corresponding contribution
to) the SC state.
\begin{figure}
\vbox{%
\centerline{\epsfig{file=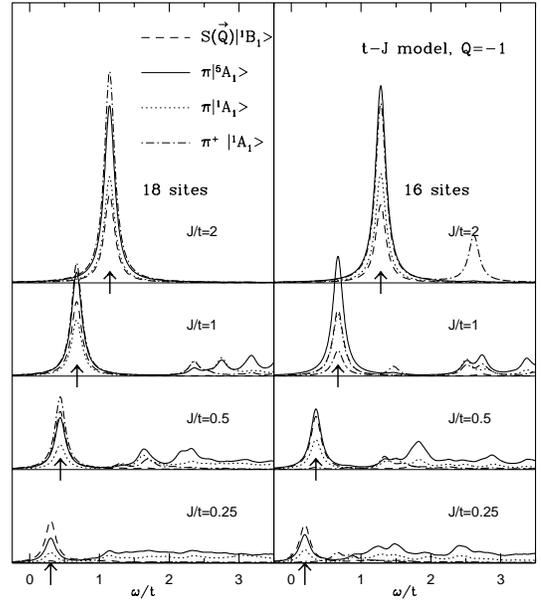,width=7.2cm}}
\narrowtext
\vspace{2mm}
\caption{\label{fig_tJ}Spectral functions with final states in the $Q$$=$$-1$
subspace: dynamical spin correlation function for momentum transfer ${\bf Q}$,
calculated for the $^1B_1(0,0)$ ground state; spectrum of the $\pi^{\dag}$
-operator, calculated for the $^1A_1(0,0)$ ground state in the $Q$$=$$-2$ 
sector;
spectra of the $\pi$-operator, calculated for the half-filled $^1A_1(0,0)$
ground state and the lowest half-filled $^5A_1(0,0)$ state.}}
\end{figure}
In Fig.~\ref{fig_tJ}, this is verified in terms of another 
``diagnostic tool'', which is the spectral function 
\begin{equation}
\label{green}
A(\omega )=\Im \frac{1}{\pi }\langle \psi |\widehat{O}^{\dag}\frac{1}{\omega
-(H-E_{ref})-i0_{+}}\widehat{O}|\psi \rangle.
\end{equation}
We see that, if we apply as the operator $\hat{O}$ the $\hat{\pi} ^{\dag}$ 
operator to the ``dSC'' state in the
$\nu =2$ irreps (the $Q=-2,S_{z}=0$ state $^{1}A_{1}(0,0)$), we end up in
precisely the same final state (to within computer accuracy) as if we apply
the $\hat{\pi} $ operator to the ``AF'' state 
$(Q=0,S_{z}=0;$ $^{5}A_{1}(0,0)$).
In other words, two successive $\hat{\pi} -$rotations rotate us from the 
AF state
to the d-SC state. The energy required to perform this rotation, $\omega _{0}$,
is again seen to scale with $J$ (in contrast to a false argument in 
ref.~\cite{Greiter}).

Finally, we note that the dynamical spin correlation function 
 -- where we take as the 
operator $\hat{O}$ in (\ref{green}) the magnon operator
 ${\bf S} ( {\bf Q} ) $ -- has a peak at
precisely the same position as the $\pi -$resonance. This confirms the
earlier conjecture by Demler and Zhang \cite{DemlerZhang} to interpret a 
well-known spin
resonance detected in neutron-scattering experiments in HTSL, as a ``$\pi -$%
excitation''. More generally, this fingerprint of SO(5) symmetry can be
described as follows: Away from $\mu _{c,}$ the chemical potential, which
controls the changes in the total charge, breaks SO(5) symmetry. As
already mentioned, the effect of this chemical potential is formally
analogous to the effect of the ${\bf B}-$field on a spin. In the presence of 
this
symmetry-breaking field, the order parameter is forced to precess and this
explicit breaking of the SO(5) invariance induces transitions between AF
and SC states, and governs the competition between these two phenomena. The 
SO(5) theory -- in good accord with the experimental data \cite{Mook}
both concerning doping and temperature dependence -- then
identifies this precession frequency $\omega _{0}$ with a resonance in
neutron scattering.

\section{SO(5) symmetry in ladders}
\label{ladders}

\subsection{Introduction}
High-$T_c$ materials are antiferromagnetic Mott insulators displaying 
long-range AF order at
half filling. The antiferromagnetic phase is rapidly destroyed upon doping
and is replaced by the superconducting phase. Below optimal doping (the
doping with maximum $T_c$) and above the superconducting temperature there
are clear experimental indications for the opening of a spin gap
\cite{BatloggEmery,Ong}.
This phase is termed spin-liquid with properties which are quite difficult
to reproduce on the
theoretical level for a two-dimensional system. On the other
hand, at and close to half filling this spin-gap variety of Mott 
insulators is obtained quite
well in  half-filled ladder systems: Upon doping the ladder, 
the spin gap persists
and the system exhibits quasi-long-range d-wave superconducting
correlations, which become dominating in some parameter range 
\cite{DagottoRice}.
In this sense, ladders systems show  properties very similar to the
phase diagram of
cuprate materials, the main difference being
the fact that correlations are ``quasi-long-ranged'', i.e.~they show
power-law behavior, since they cannot be truly ``long-ranged'' because of 
one-dimensionality. Moreover, there also exist 
copper-oxides with  CuO$_2$ planes 
containing line defects, which result in ladder-like arrangements of
Cu-atoms \cite{AzumaHiroiTakano}. These systems 
can be described in terms of coupled two-leg ladders 
exhibiting a spin gap 
and thus belong to the spin-liquid Mott-insulator variety. Also the related 
``stripe 
phases'' of the 2D CuO$_2$ planes in the cuprate
superconductors\cite{Tranquada}, which have recently received 
considerable attention, can be mapped onto ladder systems.

It is thus interesting to study the occurrence of SO(5) symmetry in
ladder systems. As we will discuss, there is in fact a natural way to
construct an SO(5) symmetric model for a two-leg ladder, which has only
local interactions on a rung of the ladder\cite{ScZhHa}. We can thus
use the ladder system as a theoretical laboratory to check some ideas
of the SO(5) theory. 
On the other hand, we want to study whether {\it generic } models, which
are
 {\it not} SO(5) invariant at the bare starting (microscopic) level may
 show SO(5) symmetry in their low-energy regime. An appropriate tool to
 study this low-energy regime, starting from a microscopic Hamiltonian,
 is the renormalization group (RG). This method has proven to be
 particularly suited  to study systems between one and two
 dimensions, for example  ladder systems \cite{Fabrizio,Schulz,BalentsFisher}.

\begin{figure}
\vbox{%
\centerline{\epsfig{file=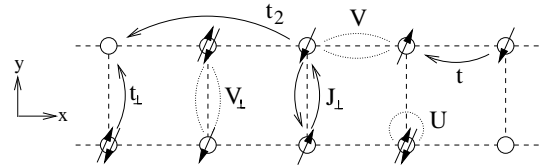,width=7cm}}
\narrowtext
\vspace{1mm}
\caption{\label{fig_ladder}A generalized Hubbard-type ladder model: 
the Hamiltonian consists of 
electron hopping terms ($t$), on-site and off-site 
Coulomb interactions ($U$, $V$), and spin-spin exchange terms ($J$)}}
\end{figure}

In the following, we consider generalized Hubbard-type ladders with 
Hamiltonians of the form
\begin{eqnarray}
\label{extHubHami}
H & = & - t\sum_{\langle i,j\rangle,\g} \left(  c_{i,\g}^{\dag} c_{j,\g} + 
          \mbox{h.c.} \right) 
        - t_\perp \sum_{i,\g}  c_{i,\g}^{\dag} c_{i \pm \hat{y},\g}\nonumber\\
  &   & + U \sum_{i}   n_{i,\up} n_{i,\down} 
        + V \sum_{\langle i,j\rangle}  n_{i} n_{j}
        + \frac{V_\perp}2 \sum_{i}  n_{i} n_{i \pm \hat{y}}\nonumber\\
  &   & + \frac{J_\perp}2 \sum_{i}  {\bf S}_{i} {\bf S}_{i \pm \hat{y}}\;.
\end{eqnarray}
(see Figure \ref{fig_ladder}).
The sum over $i$ in (\ref{extHubHami}) runs over the sites  of a two-chain
ladder, as before,
$\langle i,j \rangle$ covers pairs of nearest-neighbor sites on the same chain
and $i\pm \hat{y}$ is the nearest-neighbor site to $i$ on the
other chain.

\subsection{Exact SO(5) ladder}
In this paragraph we present ``numerical spectroscopy experiments'' for a 
recently
proposed\cite{ScZhHa} exactly SO(5) symmetric ladder model based
on a simplified version of the more general Hamiltionian (\ref{extHubHami}):
The non-local Coulomb and spin-spin interactions (the $V$ and $J$ terms in 
(\ref{extHubHami})) 
are restricted to act within the rungs, but not along the legs of the ladder,
and hopping is only allowed between nearest-neighbor sites (with hopping
$t$ along the legs and $t_{\perp}$ within the rungs) 
(see Fig.~\ref{fig_ladder}).
Setting $J_{\perp}=4(U+V_{\perp})$, the elementary magnetic and 
``SC'' excitations of a half-filled
single rung, namely the formation of a triplet and the formation of an 
electron (or hole) pair, become
degenerate in energy. In this case, one obtains an exactly SO(5) 
symmetric Hamiltonian \cite{ScZhHa}.
In a recent work \cite{EdDoZaHaZh}, this ladder model was 
studied in detail and related to the triplet-excitation picture and 
coherent-state 
description of section \ref{one_to_one}. 
Furthermore, this work gives the most general 
construction of the SO(5) irreducible representations not only 
for even numbers of
holes (or electrons) as in section \ref{micro_models}, 
but also for odd numbers,
encountered for example in direct or inverse photoemission (PE). 
Exact diagonalization
was used to extract the practical application
of the selection rules for photoemission implied by SO(5) symmetry.
\begin{figure}[bth]
\vbox{%
\centerline{\epsfig{file=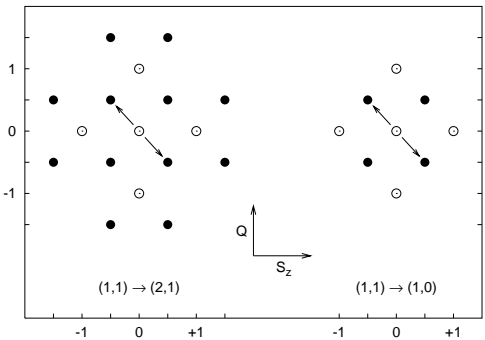,width=6.5cm}}
\narrowtext
\vspace{1mm}
\caption{\label{fig_mult1}SO(5) multiplets connected by allowed photoemission 
(PES) and inverse photoemission (IPES) transitions. 
As we inject/remove a $\down$-electron,
PES corresponds to an arrow pointing south-east and IPES to one pointing 
north-west. 
Here, the initial state is a half-filled state with $S_z$$=$$0$ in the $(1,1)$ 
multiplet (open circles). SO(5) selection rules only allow transitions to the 
$(2,1)$ and the $(1,0)$ multiplets (black circles), which are superimposed 
to the initial $(1,1)$-multiplet.}}
\end{figure}

\begin{figure}[bth]
\vbox{%
\centerline{\epsfig{file=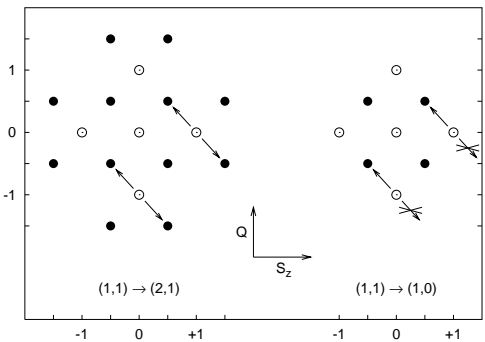,width=6.5cm}}
\narrowtext
\vspace{1mm}
\caption{\label{fig_mult2} 
PES/IPES from a spin-polarized 
half-filled state ($Q$$=$$0$, $S_z$$=$$+1$) and from a doped state 
($Q$$=$$-1$,$S_z$$=$$0$). Both states are members of the 
$(1,1)$ multiplet like the initial 
state in Fig.~\ref{fig_mult1}. Note that there is no allowed photoemission 
transition to the $(1,0)$ multiplet for either of the two states.}}
\end{figure}

The classification of SO(3) symmetric states into spin multiplets, e.g.~a
triplet with $S$$=$$1$, is well known. For SO(5)
symmetry a similar classification holds, but to characterize a multiplet
one needs two integer quantum numbers $(p,q)$ \cite{EdHaZh,EdDoZaHaZh}:
within a given $(p,q)$ multiplet the states are characterized both by $S_z$ 
and the charge $Q$ (instead of only $S_z$ in SO(3)).  
Fig.~\ref{fig_mult1} and \ref{fig_mult2} display some of the lowest SO(5) multiplets
as well as some allowed PE transitions between them (note that for $p$$=$$q$, 
one gets the diamond-like structures shown in Fig.~\ref{fig_irreps}).

These theoretical assertions can be checked numerically using the Lanczos 
technique
\cite{Dagotto}. The method permits to calculate
ground state properties like energy and spin expectation values, 
but also Green's 
functions (cf.~(\ref{green}) ), and, in particular, the photoemission and 
inverse photoemission spectra (here for a spin-down electron):
\begin{eqnarray*}
A_{PES}({\bf k},\omega) &=& \frac{1}{\pi} \Im
\langle 0 |c_{{\bf k}\downarrow }^\dagger\frac{1}{
\omega +H-\epsilon_0-i0^+}c_{{\bf k}\downarrow }| 0 \rangle,
\nonumber \\
A_{IPES}({\bf k},\omega) &=& \frac{1}{\pi} \Im
\langle 0 |c_{{\bf k}\downarrow }\frac{1}{
\omega +H-\epsilon_0-i0^+}c_{{\bf k}\downarrow }^\dagger| 0 \rangle.
\end{eqnarray*}
$|0\rangle$ is an initial energy-eigenstate found by Lanczos-diagonalization
and $\epsilon_0$ its energy. 

With this ``computer experiment'' one can study the direct and inverse PE 
within a
multiplet structure and observe crucial selection rules. The SO(5) multiplets 
are easily identified, because the energies of states belonging to one 
multiplet
are degenerate (i.e.~identical to computer accuracy of ca.~$10^{-14}$). 
This allows to study the evolution of the single-particle spectral function
as we pass from one multiplet ($p,q$) to the other, and 
within one multiplet through the different doping and spin levels $(Q,S_z)$.

\begin{figure}[h]
\vbox{%
\centerline{\epsfig{file=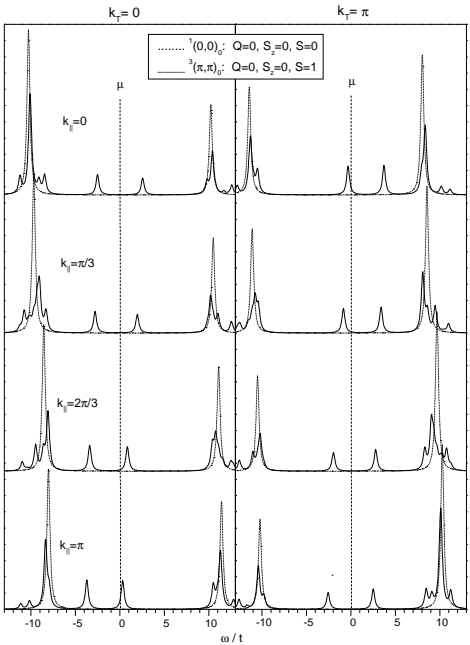,width=6.4cm}}
\narrowtext
\vspace{2mm}
\caption{\label{fig_pesipes1}PES/IPES spectra of a $6$-rung ladder 
(removal/injection of a $\down$-electron. $\mu$ is the Fermi energy,
defined as the average of first ionization and affinity energy.
Parameter values are $U/t$$=$$8$, $V_{\perp}/t$$=$$-6$, $J_{\perp}/t$$=$$8$, 
$t_{\perp}/t$$=$$1$.
The dotted line shows PES/IPES from the half-filled ground state 
($S_{tot}$$=$$0$), the full line from the half-filled $^3(\pi,\pi)_0$ 
state with $S_z$$=$$0$ and $S_{tot}$$=$$1$ (cf.~Fig.~\ref{fig_mult1})}}
\end{figure}

The dotted line in Fig.~\ref{fig_pesipes1} shows the single-particle 
spectrum
for the half-filled ground state $^1(0,0)_{0}$ (where the group theoretical 
notation
is a short hand for $^{degeneracy(=S)}(momentum)_{Q}$); this is an 
RVB-state 
of rung-singlets (see section \ref{intro}) which actually forms a 
one-dimensional 
$(0,0)$ multiplet. Final states can only
belong to the $4$-dimensional $(1,0)$ irreps \cite{EdDoZaHaZh}. Despite
the fact that we are using very strong interaction parameters, there
is just one single electron-like band in PES, whose cosine-type-of
dispersion closely follows the dispersion of a non-interacting electron.
The center of gravity of this band is given by the energy difference between 
a rung-singlet and a singly occupied rung \cite{EdDoZaHaZh}.

The solid line in Fig.~\ref{fig_pesipes1} represents the PES for 
the half-filled $^3(\pi,\pi)_0$ state (with $S_z$$=$$0$), 
which carries one triplet excitation (magnon) and which belongs to 
the fivefold degenerate $(p$$=$$1,q$$=$$1)$
multiplet, i.e.~the open circles in Fig.~\ref{fig_mult1}. 
From Fig.~\ref{fig_mult1} we expect in this case final states belonging 
to both the $(2,1)$ and the $(1,0)$ irreps.
Indeed, in addition to the band seen in the ground state spectra, 
which remains 
practically unchanged, there appears a ``sideband'' close to $\mu$ 
both in PES an in
IPES; thus 
\begin{figure}[h]
\vbox{%
\vspace{-1mm}
\centerline{\epsfig{file=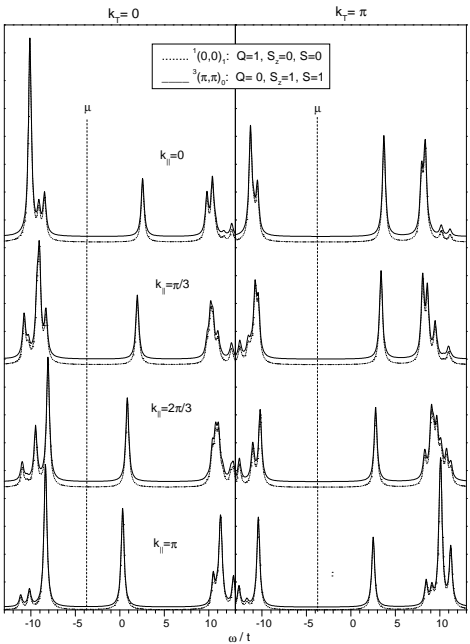,width=6.4cm}}
\narrowtext
\vspace{2mm}
\caption{\label{fig_pesipes2} 
PES/IPES from the two-hole ground state $^0(0,0)_{-1}$ (dotted line) 
and the half-filled $^3(\pi,\pi)_0$ state with $S_z$$=$$S_{tot}$$=1$ 
(full line) of a $6$-rung ladder (cf.~Fig.~\ref{fig_mult2}). 
The removed electron has $\downarrow$-spin. 
The spectra for the half-filled state have been offset in
$y$-direction so as to faciliate the comparison.
Parameter values are as in Fig.~\ref{fig_pesipes1}.}} 
\end{figure}
\noindent we have precisely the 4 bands expected from 
Fig.~\ref{fig_mult1}.
The physical interpretation is as follows: 
the initial $^3(\pi,\pi)_0$ state carries a single 
triplet-like Boson in the ``RVB-vacuum''. The ``main bands'' result from a 
creation
of the photohole in a singlet rung; then the final state, consisting 
of the initial 
triplet plus a propagating singly-occupied rung carrying the entire 
momentum transfer,
belongs to the $(2,1)$ irreps. As the two excitations, triplet and hole, 
can scatter from each other, the main band becomes broadened.
But the photohole can also be created in the initial triplet, which 
results in a 
singly-occupied rung propagating in the RVB-vacuum, thus the new state belongs
to the $(1,0)$ irreps. Since in this second case the photohole has to 
absorb the 
momentum of the initial triplet, $(\pi,\pi)$, the sidebands' dispersion is 
shifted by $(\pi,\pi)$ with respect to the main bands.

Proceeding to a $^3(\pi,\pi)_0$ initial state with $S_z$$=$$1$ 
(instead of $S_z$$=$$0$)
(Fig.~\ref{fig_pesipes2}, full line), we get a PES/IPES spectrum 
similiar to the
one for $S_z$$=$$0$, but the sideband in PES has disappeared, whereas the 
one in IPES
has gained some additional weight. The interpretation follows from 
Fig.~\ref{fig_mult2}: As the initial state belongs to the $(1,1)$ 
irreps, 
a PES transition into the $(1,0)$ multiplet is now impossible.

At this point, we remember that the SO(5) symmetry of our model 
implies that spin
polarization and hole doping are equivalent in that the empty rung 
is the ``SO(5)
partner'' of the triplet rung. Consequently, 
Fig.~\ref{fig_pesipes2} also shows
the spectra originating from the two-hole ground state $^0(0,0)_{-1}$ 
(dotted line).
Both states belong to the $(1,1)$ irreps. Accordingly, their energies agree to 
computer accuracy. 
Since also the allowed final states of these two initial states belong 
to one and
the same multiplet, namely $(2,1)$, 
we expect their PES/IPES amplitudes to be directly related.
A more detailed consideration \cite{EdDoZaHaZh} shows that their 
amplitudes are in
fact identical -- just as Fig.~\ref{fig_pesipes2} demonstrates.
The physical reason for this identity is that the photohole cannot be 
created in the
triplet rung, because the latter only contains two spin-{\it up} electrons.

In summary our numerical ``experiments'' demonstrated, that SO(5) symmetry 
implies that the single particle spectra in the doped ground state are
identical to those of certain higher-spin states at half-filling.

\subsection{Recovering SO(5) symmetry at low energies}

The renormalization-group (RG) route has been proven to be particularly suited 
to describe the low-energy behavior of systems
between one and two dimensions \cite{Fabrizio,Schulz,BalentsFisher}.
For this reason, we have carried out a RG study of  a 
rather general  ladder model of the Hubbard type\cite{ArHa}.
The main result is that the {\it effective}
Hamiltonian of  the system, i.e.~the Hamiltonian which 
describes excitations below a certain energy $\omega^*$,
becomes  SO(5) invariant  for a wide range of {\it bare} 
models. These include interactions with quite arbitrary values of 
$U$, $V$, and $V_\perp$ (cf.~Fig.~\ref{fig_ladder}), provided they are weak,
as well as a moderate next-nearest-neighbor hopping $t_2$.
It is, thus, not necessary to introduce unphysical values for the
parameters by hand in the model, since the exact SO(5) ladder
discussed in section \ref{ladders}.2 will be eventually recovered at low 
energies.
This result is remarkable, since it shows that quite general 
Hubbard-like models, which are  relevant  for the description of the cuprate
materials between 1D and 2D, although not explicitly SO(5) invariant on the 
{\it microscopic} level, display an {\it exact } SO(5) symmetry 
when observed on a {\it macroscopic} length scale $\propto
E_F/\omega^*$ (in units of the lattice spacing) \cite{ArHa}
(for the $t_2=0$ case, see also Ref.~\cite{LinBalentsFisher}).
Physically, this means that the ``standard deviation'' between the
multiplet splitting (cf.~Fig.~\ref{fig_irreps}) 
goes to zero in these models for
low energies or  large length scales.
An important issue here is the introduction of a
next-nearest-neighbor term $t_2$, which is  known to 
be essential  in order to correctly describe AF correlations and 
Fermi-surface (FS) topology in cuprate materials.

Specifically, we have considered  two coupled chains  with total 
low-energy Hamiltonian $H=H_0 + H_I$, where $H_0$ represents the
non-interacting part and $H_I$ the interaction.
This Hamiltonian describes  interacting Fermions 
(expressed by creation and destruction operators
$c_{\koms}$ and $c_{\koms}^{\dag}$)
close to the FS. The FS 
  consists of four points: two
bands $\ky=0$ and $\ky=\pi$, each one with two Fermi points
corresponding to
right- and left-moving
Fermions. 
Since we are interested in low-energy properties very close
to the FS, the Fermion dispersion can safely be taken as linear
around the FS 
 with (in general) band-dependent velocities $v_F$.

The idea of the RG is to divide the electronic excitations within the
Brillouin zone into 
high-energy and low-energy modes, the latter being
modes  restricted within  an energy $\om\ll E_F$ from  the Fermi energy. 
One then eliminates  the high-energy modes by integrating them out
and constructs an effective Hamiltonian which is
restricted to the low-energy excitations. 
The parameters of the new effective low-energy Hamiltonian thus depend on the
energy cutoff $\om$.
Also the total spectral weights at the FS (quasiparticle weights) 
$\zz{\ky}{\lflow}$ 
of the two bands 
 are reduced due to the reduction of the cutoff  and depend on $\lflow$.
 For this reason, in order to
recover the {\it canonical} Fermi operators within the low-energy subspace, 
one should reabsorb $\zz{\ky}{\lflow}$ into the definition of the 
Fermi operators and transform to
$ c_{\koms}=   \sqrt{\zz{\ky}{\lflow}}  \wtilde c_{\koms}$.
This transformation has to be done in order to preserve the sum rules for 
the total integrated
spectral weight $Z=1$ within the restricted subspace.
The $\wtilde c_{\koms}$ now acquire the meaning of {\it canonical}
operators with the correct anticommutation
relations within the low-energy subspace.

In practice, the integration of the high-energy modes is carried out 
by decreasing $\lflow$ via
infinitesimal steps, starting at $\lflow= E_F$ which corresponds to the bare
(microscopic) Hamiltonian (see, e.g.~\cite{soly.79}).
Due to the restriction of the modes to a small ``shell'' around the FS,
the scattering amplitudes in the interaction $H_I$
can also be considered as 
dependent only on the Fermi momenta closest to where the corresponding
 processes take  place as well as on the spin of the scattered
 particles \cite{soly.79,Fabrizio,BalentsFisher}. 

The  SO(5)invariant part $H_I^{(\sofm)}$ of $H_I$ 
 can conveniently be written 
as a sum of products
of SO(5) scalar contractions in the form 
$\Psi^{\dag}_{\kom_1}  \Psi_{\kom_2} 
 \times 
\Psi^{\dag}_{\kom_3}  
 \Psi_{\kom_4} $ 
where $ \Psi_{\kom}$ are
SO(5) four-spinor Fermi operator  
(see Refs.~\cite{ra.ko.97,ArHa} for details).
At $\lflow\approx E_F$, the bare Hamiltonian is {\it not} SO(5)
 invariant in general and
$H_I$ consists of two parts 
$H_I = H_I^{(\sofm)} + H_I^{({\rm breaking})}$, where $H_I^{({\rm
    breaking})}$ is a symmetry-breaking term.
On the other hand, for $t_2=0$, 
the non-interacting part $H_0$ is  $\sof$ symmetric, whereas a
finite  $t_2$  breaks PH (and thus SO(5)) symmetry 
also in $H_0$ 
through a difference
$\dvfn$ between the Fermi velocities of the two bands.
 
\begin{figure}
\vbox{%
\centerline{\epsfig{file=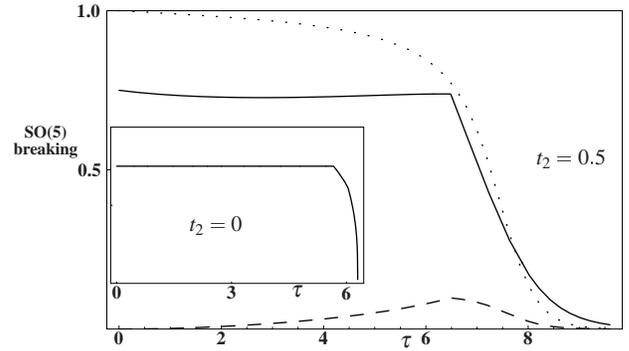,width=8.3cm}} 
\narrowtext
\vspace{2mm}
\caption{\label{figproc}
RG flow of the SO(5)-breaking terms of the Hamiltonian (in arbitrary units)
$H_I^{({\rm breaking})}$ (full), $H_I^{({\rm PH-breaking})}$ (dashed),
$\dvfn$ (dotted)
 as functions of the RG flow $\tau=-\log\lflow/E_F=$ for  $U=1.$,
$t=\tp=1$, $t_2=0.5$ and half filling.
The inset shows  the $t_2=0$ case where only 
$H_I^{({\rm breaking})}$ is nonvanishing.}}
\end{figure}

The striking result of our calculation is that, by integrating away the 
high-energy modes, from some energy scale $\om=\omega^*$ on,
the SO(5)-breaking part of the Hamiltonian $ H_I^{({\rm breaking})}$ 
eventually vanishes with respect to the SO(5)symmetric part.  
In the inset of Fig.~\ref{figproc}, we show
the RG flow for the $t_2=0$ case (here the SO(5)-breaking term appears
only in the interacting part of the Hamiltonian). In the bulk of the
same figure, the flow for the $t_2\not=0$ case 
is displayed, for which there are
three types of SO(5)-breaking terms, namely, $\dvfn$, a
PH-breaking term $H_I^{({\rm PH-breaking})}$ and a plain SO(5)-breaking
term $H_I^{({\rm breaking})}$. All three terms go to zero at $\lflow^*$.

This result is of extreme importance, since it means 
that even though the system is  not SO(5) invariant at the {\it bare} level,
i.~e.~at $\lflow=E_F$,
the SO(5) invariant part of the Hamiltonian eventually dominates with
respect to the symmetry-breaking part at  energies smaller than $\omega^*$.
We have checked that this  occurs 
for very general values of the Hamiltonian, including on-site ($U$) and
nearest-neighbor  ($V$ and $V_\perp$) interactions, provided they are weak.

An interesting result is that for the $t_2\not=0$ case, the
quasiparticle weights 
$\zz{\ky}{\lflow}$ for the two bands renormalize differently.
These quasiparticle weights have to be reabsorbed into the definition of the
canonical Fermi operators through the transformation to the 
$ \wtilde c_{\koms}$ variables, as explained above.
This has the important consequence 
that the new {\it low-energy} SO(5) invariant Hamiltonian 
$H_I^{({\sofm})}$ is now invariant under a {\it renormalized} 
SO(5) symmetry in terms of {\it new} $\hat{\pi}$ operators
(cf.~(\ref{def_pi+}) ), whereby the Fermi operators 
$ c_{\koms}$ are replaced with $\wtilde c_{\koms}$.
This  extended  concept of SO(5) symmetry makes it possible for this
{\it generalized} symmetry to occur
in a {\it larger} and more {\it generic} class of physical systems than the 
ordinary SO(5). Moreover, a SO(5) theory unifying antiferromagnetism
and superconductivity in terms of this {\it generalized}
representation can admit possible asymmetries between the
antiferromagnetic and superconducting phase, like for example the
difference in $T_c$ or in the order parameter\cite{Zhang,henl.97}.

In conclusion, the renormalization-group study shows that the effective 
low-energy Hamiltonian of a 
quite generic Hubbard-like ladder with weak interaction is SO(5) symmetric.
This holds true also with the inclusion of a next-nearest-neighbor hopping  
$t_2$, provided it is 
{\it written in terms of the appropriate canonical Fermi operators
 $\wtilde c$ for the low-energy subspace}.
\medskip

{\it Acknowledgements}: The authors would like to express their deep
gratitude to S.~C.~Zhang for sharing his deep insights and enthusiasm 
on SO(5) theory and D.~J.~Scalapino for a very fruitful interaction on the
exact SO(5) ladder.


\end{multicols}
\end{document}